# Remote Interactions between tropical cyclones: The case of Hurricane Michael and Leslie's high predictability uncertainty

M. López-Reyes [a,b,c,*], J.J. Gónzalez-Alemán [d], C. Calvo-Sancho [e], P. Bolgiani [b], M. Sastre [b], M.L. Martín [e,f]

[a] Astronomy and Meteorology Institute (IAM), Physics Department, University of Guadalajara, Guadalajara, Mexico
[b] Department of Earth Physics and Astrophysics, Faculty of Physics, Complutense University of Madrid, Madrid, Spain
[c] Instituto Frontera A.C., Tijuana, México
[d] Agencia Estatal de Meteorología (AEMET), Department of Development and Applications, Madrid, Spain
[e] Department of Applied Mathematics, Faculty of Computer Engineering, University of Valladolid, Spain
[f] Institute of Interdisciplinary Mathematics (IMI), Complutense University of Madrid, Madrid, Spain

ABSTRACT

The study explores Hurricane Michael's impact on Hurricane Leslie's trajectory predictability using ECMWF and NCEP ensemble systems. A clustering method focused on tropical cyclones is used to identify potential paths for Leslie: Cluster 1 accurately predicted Leslie's direction towards the Iberian Peninsula, whereas Clusters 2 and 3 indicated a southern recurve near the Canary Islands. Analysis of potential vorticity and irrotational wind at upper levels showed a significant interaction between Michael, ridge, and trough across the jet stream from +12 h after initialization. Cluster 1 showed a stronger Michael promoting upper-level wind divergence greatest, modifying the jet stream configuration around the ridge and downstream. Alterations in the jet stream's configuration, functioning as a waveguide, propagated downstream, guiding Leslie towards the Iberian Peninsula. Clusters 2 and 3 indicated the trough's failure to incorporate Leslie, resulting in a recurve of the trajectory around the Azores anticyclone. This research enhances comprehension of the interaction between two tropical cyclones via synoptic Rossby wave flow. Moreover, the conceptual framework can aid operational meteorologists in identifying the sources of uncertainty, particularly in track forecasts under synoptic conditions analogous to those examined in this study.

## 1. Introduction

Accurate trajectory forecasts of tropical and extratropical cyclones are crucial for decision-making and minimizing direct impacts on the population and economic activities (Morrow and Lazo, 2015). In the last decade, forecasts of tropical cyclone (TC) tracks have experienced very notable improvements (Heming et al., 2019). However, the uncertainty in the trajectory forecasts is noticeably higher under certain atmospheric scenarios, e.g., during extratropical or tropical transitions, as well as in events where two cyclones very close to each other interact (Grams et al., 2018; Heming et al., 2019; Sánchez et al., 2020).

In extratropical latitudes, the trajectories of various cyclone types in the synoptic scale, including tropical and extratropical cyclones, are influenced by the wind fields in atmospheric mid and upper-levels (Grams et al., 2015; Keller et al., 2019; Zhang et al., 2019). The behavior of Rossby waves is a condition for the development of different meteorological phenomena in extratropical latitudes associated with cyclone and anticyclone circulations in different vertical levels (Hoskins et al., 1985). Moreover, Rossby waves act as precursors to high-impact weather events (Berman and Torn, 2019; Chaboureau and Claud, 2006) such is the case of hurricane Leslie's extratropical transition (ET; Mandement and Caumont, 2021; López-Reyes et al., 2023). Several studies (Keller et al., 2019; Sánchez et al., 2020; Weijenborg and Spengler, 2020) have suggested that downstream forecast errors are associated with upstream diabatic heating processes where the upper-flow divergence stimulates the extension of the ridge. This amplified ridge can propagate by amplifying the ridge-trough system. Anwender et al. (2008), Riemer and Jones (2010), Archambault et al. (2013) and Berman and Torn (2019) studied other mechanisms that contribute to an upstream ridge amplification, such as adiabatic transport of potential vorticity (*PV*), adiabatic reduction above the maximum diabatic heating level as well as air advection by divergent flows at tropospheric upper-levels. Consistently, previous works have suggested that the amplitude of ridges is influenced by remote sources of irrotational winds, which are driven by diabatic processes linked to TCs. This relationship holds true even in the context of extratropical transition (ET) processes characterized by highly convective environments (Torn et al., 2015; Riemer et al., 2008).

* Corresponding author at: Department of Earth Physics and Astrophysics, Faculty of Physics, Complutense University of Madrid, Madrid, Spain.
*E-mail address:* maurilop@ucm.es (M. López-Reyes).

ET is the process by which a TC loses its tropical structure and develops characteristics typical of extratropical cyclones (Bieli et al., 2019). During ET, simultaneous thermodynamic and dynamic processes occur across different atmospheric scales. This typically happens as a TC moves into higher latitudes and encounters a baroclinic environment characterized by cooler sea surface temperatures, meridional humidity gradients, strong vertical wind shear, and an increased Coriolis parameter (Jones et al., 2003; Evans et al., 2017; Bieli et al., 2019). One of the most favorable conditions that promote ET occurs when a TC interacts with an upper tropospheric trough in a region of high wind divergence (Jones et al., 2003).

In October 2018, during the final phase of Hurricane Leslie life cycle, as it underwent an ET, the numerical prediction models struggled to accurately forecast its trajectory and intensity, even within a few hours of model initialization (López-Reyes et al., 2023). This challenge was particularly evident in the Ensemble Prediction System (EPS) from the European Center for Medium-Range Weather (ECMWF) where a 24-h forecast displayed high uncertainty in Leslie's path (López-Reyes et al., 2023). Some trajectories projected Leslie towards the Iberian Peninsula, while others showed a southward recurve, bringing the system very close to the Canary Islands. Remarkably, during this time framework of heightened uncertainty in Leslie's trajectory forecasts, Hurricane Michael was undergoing an ET process upstream (Beven II et al., 2019). Previous studies (Sánchez et al., 2020; Weijenborg and Spengler, 2020) have suggested that diabatic processes occurring upstream can induce alterations in the behavior of Rossby waves. Such modifications may be a contributing factor to the escalating Leslie trajectory uncertainty downstream (Teubler and Riemer, 2016; Grams et al., 2015; Keller et al., 2019; Sánchez et al., 2020). The intricate interplay between Leslie's transition and the influence of Hurricane Michael upstream adds a layer of complexity to our understanding of the atmospheric dynamics during this critical phase.

Under atmospheric conditions characterized by high uncertainty, particularly in the trajectory and intensity forecasts of TCs, ensemble prediction systems offer valuable insights into the event's predictability (Sarkar et al., 2021; Sattar et al., 2023; Morss et al., 2024). These systems help identify key sources of uncertainty, assist forecasters in refining their interpretations, and enhance the accuracy of predictions compared to deterministic models, thereby offering a more robust probabilistic forecast (Lu et al., 2024). Motivated by the case of the high uncertainty in the track and intensity forecast of the hurricane Leslie, the purpose of this study is to explore the possible influence of Hurricane Michael upstream and its downstream impact on the predictability of Leslie. To do this, several trajectories of Hurricane Leslie are clustered from the operational ECMWF outputs of EPS and the Ensemble Forecast System (GEFS) from National Center for Environmental Prediction (NCEP) are selected building a superensemble during the period of Leslie's highest uncertainty in the forecast.

This work is organized as follows: the methodology used is presented in Section 2; the results and discussions are shown in Section 3 and finally, summary and conclusions are found in Section 4.

## 2. Materials and methods

### 2.1. Dataset

Forecast data for this study are from THORPEX The Interactive Grand Global Ensemble (TIGGE) database at ECMWF (ECMWF, https://confluence.ecmwf.int/display/TIGGE) similar to Kowaleski and Evans (2016). We employ perturbed forecasts ensemble generated by two of the main global models: on the one hand, the Integrated Forecasting System from ECMWF (IFS, Cycle 41r2: European Center for Medium-Range Weather Forecasts, 2023) with 0.5° × 0.5° of horizontal resolution, 137 vertical levels and 50 perturbed members, and on the other hand, the Global Forecast System from NCEP (GFS, v14.0.0: National Center Environment Information, 2023), with 0.5° × 0.5° of horizontal resolution, 64 vertical levels and 20 perturbed members.

Forecasts are analyzed starting with initialization on October 8, 2018, at 0000UTC and using initializations every 12 h until October 12, 2018, at 1200UTC; the time steps in each initialization are 6 h. To investigate the possible influence of Hurricane Michael on the trajectory uncertainty of Hurricane Leslie's ET, we selected the last initialization in which both models show the highest spread. This initialization corresponds to October 11 at 0000 UTC. Mean sea level pressure (MSLP), *PV* at 320 K isentropic level, zonal ($u$) and meridional ($v$) wind components and geopotential height ($Z_{300}$) are used at 300 hPa.

An algorithm based on MSLP minimum is used to track the trajectory of members of each ensemble for the first 120 h after each simulation initialization. We define superensemble as the EPS and GEFS members combination. Mean standard deviation (MSTD; Eq. (1)) is used for measuring the dispersion of all members respect to superensemble mean. In this case, MSTD is the averaged STD (Eq. (2)) during Leslie's life cycle.

$$MSTD = \frac{\sum_{i=1}^{t_n} STD_i}{t_n} \quad (1)$$

where

$$STD = \sqrt{\frac{\sum_{i=1}^{n} (x_i - \overline{x})^2}{n}} \quad (2)$$

$t_n$ is a number of time steps and $n$ is a number of members.

As in the tracking algorithm, MSLP is the key variable to calculate Leslie and Michael's speed for each member. Following numerous studies (Hoskins et al., 1985; Teubler and Riemer, 2016; Berman and Torn, 2019), the *PV* at 330 K isentropic level ($PV_\theta$) is used as a tracer of the synoptic flow, under an adiabatic environment. The Ertel Potential Vorticity (Eq. (3)) associates the potential temperature field with the $PV_\theta$ is

$$PV_\theta = \frac{g}{\rho}(f + \xi_\theta)\nabla\theta \quad (3)$$

where, $g$ is a gravitational acceleration, $\rho$ is the air density, $f$ is the planetary vorticity, $\xi_\theta = (\partial v/\partial x - \partial u/\partial y)_\theta$ is the isentropic relative vorticity and $\theta$ is the potential temperature. However, to explain the adjustments in isentropes due to forcing caused by the release of latent heat in convective processes, such as in Hurricane Michael, the Lagrangian potential vorticity trend equation is used (Hoskins et al., 1985; Calvo-Sancho et al., 2022)

$$\rho\frac{dPV_\theta}{dt} = (f + \xi_\theta)\nabla\dot{\theta} \quad (4)$$

where $\dot{\theta}$ is the diabatic heating rate (i.e., $d\theta/dt$).

### 2.2. Clustering methodology

Following Kowaleski and Evans (2016) clustering technique, we analyze the forecast track uncertainty. During the Leslie ET phase, all tracks of the superensemble are grouped, highlighting differences in the upstream (downstream) pattern where Michael (Leslie) is located. The clustering method is based on a finite mixture model (Everitt and Hand, 1981; Camargo et al., 2007; Gaffney et al., 2007) and is calculated from coordinates (latitude and longitude) with time steps every 6 h over a total simulation period of 168 h. The Finite mixture model technique employs a mixture polynomial regression model to fit the shape of trajectories (Camargo et al., 2007; Gaffney and Smyth, 1999; Gaffney, 2004), i.e., the shapes of the trajectories are adjusted using a polynomial function. This approach demonstrates the advantage of modeling non-Gaussian distributions (Kowaleski and Evans, 2016), as well as the ability to cluster parameters that vary over time, such as the trajectories of TC. This clustering method requires specifying the polynomial order and number of clusters prior to clustering.

To determine the optimal number of clusters and the optimal polynomial order in this study, longitudinal and latitudinal position versus time with Bayesian information criterion (BIC; Eq. (5)) along with visual inspection are here considered (Camargo et al., 2007; Kowaleski et al., 2020),

$$BIC(n) = k\,ln(n) - 2ln(L) \tag{5}$$

where $k$ is the number of parameters estimated by the model, and $L$ is the maximized value of the log-likelihood function of the model defined as $L(\theta|X) = \prod_{i=1}^{n} P(x_i|\theta)$, where $\theta$ are the gaussian model parameters, and $x_i$ are the individual observations (latitude and longitude). Unlike the $L(\theta|X)$ favors more complex models, the BIC prioritizes models that strike a balance between goodness-of-fit and simplicity (Don et al., 2016). The optimal polynomial order and clusters number are determinate for the curve with lower BIC values and when the changes in the slope of the curve are low, respectively. A more detailed description of this clustering method is presented in Gaffney (2004). Evans et al. (2017) and González-Alemán et al. (2018) describe the use of this clustering method applied in TCs.

### 2.3. Composite analysis

From the three different clusters, composites of $PV_\theta$, MSLP, $Z_{300}$ and irrotational wind magnitude at 300 hPa ($\left|\overrightarrow{V}_{irr}\right| = u_{irr}^2 + v_{irr}^2$) based on Helmholtz decomposition (Chorin et al., 1990) are calculated at every time step, where $\overrightarrow{V}_{irr}$ is the irrotational wind vector and $(u_{irr}, v_{irr})$ are their components. According with Helmholtz theorem, the horizontal wind $\overrightarrow{v} = (u,v)$, can be represented by stream function ($\psi$) and potential velocity ($\chi$) as

$$\overrightarrow{v} = \widehat{k} \times \nabla\psi + \nabla\chi.$$

Here, $\widehat{k}$ is the unit vector in vertical direction and $\nabla$ is the gradient operator in two dimensions. The second term of the equation is so called irrotational or divergent wind (details in Chorin et al., 1990 and Cao et al., 2014).

To identify changes in synoptic patterns, differences between the clusters are calculated. The non-parametric Mann-Whitney-Wilcoxon test is used with 0.05 statistical significance for the composite differences between clusters. Additionally, average of $\left|\overrightarrow{V}_{irr}\right|$ is calculated for each cluster within a 500 km radius around the center of Michael and represented using box and whisker plots. The previous procedure is repeated for the most extreme members of clusters 1 and 2. Specifically, the members corresponding to the 80th percentile of cluster 1 ($P_{80}$ cluster 1) and the 20th percentile of cluster 2 ($P_{20}$ cluster 2) are considered. Finally, to identify the relations between Michael intensity and upper-level irrotational wind, the MSLP of the Michael system is calculated for each cluster with time steps of 6 h.

## 3. Results and discussions

### 3.1. Superensemble clustering analysis of Leslie's trajectories

On October 11, 2018, Leslie was very close to being embedded in the mid-latitude flow, located southeastern of a trough, near the region of maximum upper-level wind divergence (Pasch and Roberts, 2019), being its ET favored by local dynamics (Dacre and Gray, 2013; Fig. S1 in Supporting Information S1). Simultaneously, the major Hurricane Michael is upstream rapidly moving north-northeastern where it begins to interact with the trough-ridge off the USA eastern coast (Beven II et al., 2019; Fig. S1 in Supporting Information S1).

Following the BIC method (Eq. (5)), the optimal number of clusters for Leslie trajectory is three, based on third-order polynomials (Fig. 1a). The initialization on October 11 at 0000 UTC is the last one that shows high dispersion among the superensemble members (Eq. (1); shaded region in Fig. 1b). In subsequent initializations, the uncertainty in the trajectories greatly decreases (Fig. 1b). For this reason, the initialization is set on October 11 at 0000 UTC.

Two main types of trajectories are identified on this date: on the one hand, the ones following an eastern-northeastern movement (cluster 1 in Fig. 2a, heading towards the Iberian Peninsula) and on the other hand, the ones that recurve towards the south, in proximity to the Canary Islands (clusters 2 and 3 in Fig. 2a), associated with the circulation around the Azores anticyclone. All trajectory clusters show a high dispersion with respect to the members of each cluster, especially at +38 h lead time (bars in Fig. 2b). Errors in cluster 1 are notably smaller than those of clusters 2 and 3, with respect to the best track (Beven II et al., 2019; Fig. 2a and b). To clearly differentiate and identify the atmospheric patterns related to the upstream influence/interaction of Hurricane Michael in the development and displacement of Leslie, here only the extreme clusters (clusters 1 and 2 in Fig. 2a) are considered. For the whole study, we consider cluster 1 as a reference due to its similarity with the best track, regarding the mean value.

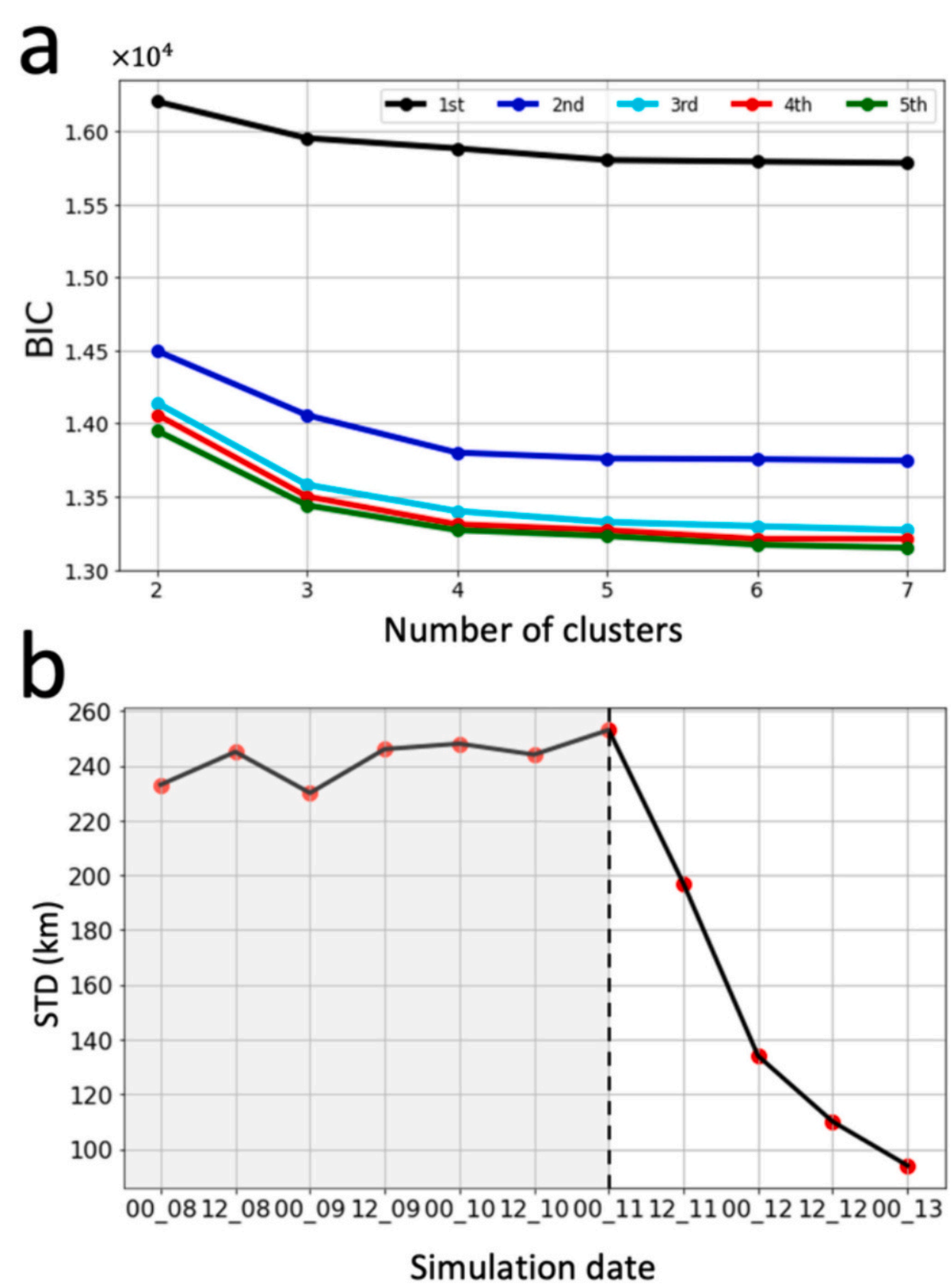

**Fig. 1.** (a) BIC for regression mixture model used in superensemble clustering hurricane Leslie at 0000 UTC October 11; (b) STD of superensemble members for different initializations from 0000 UTC October 8 to 0000 UTC October 13.

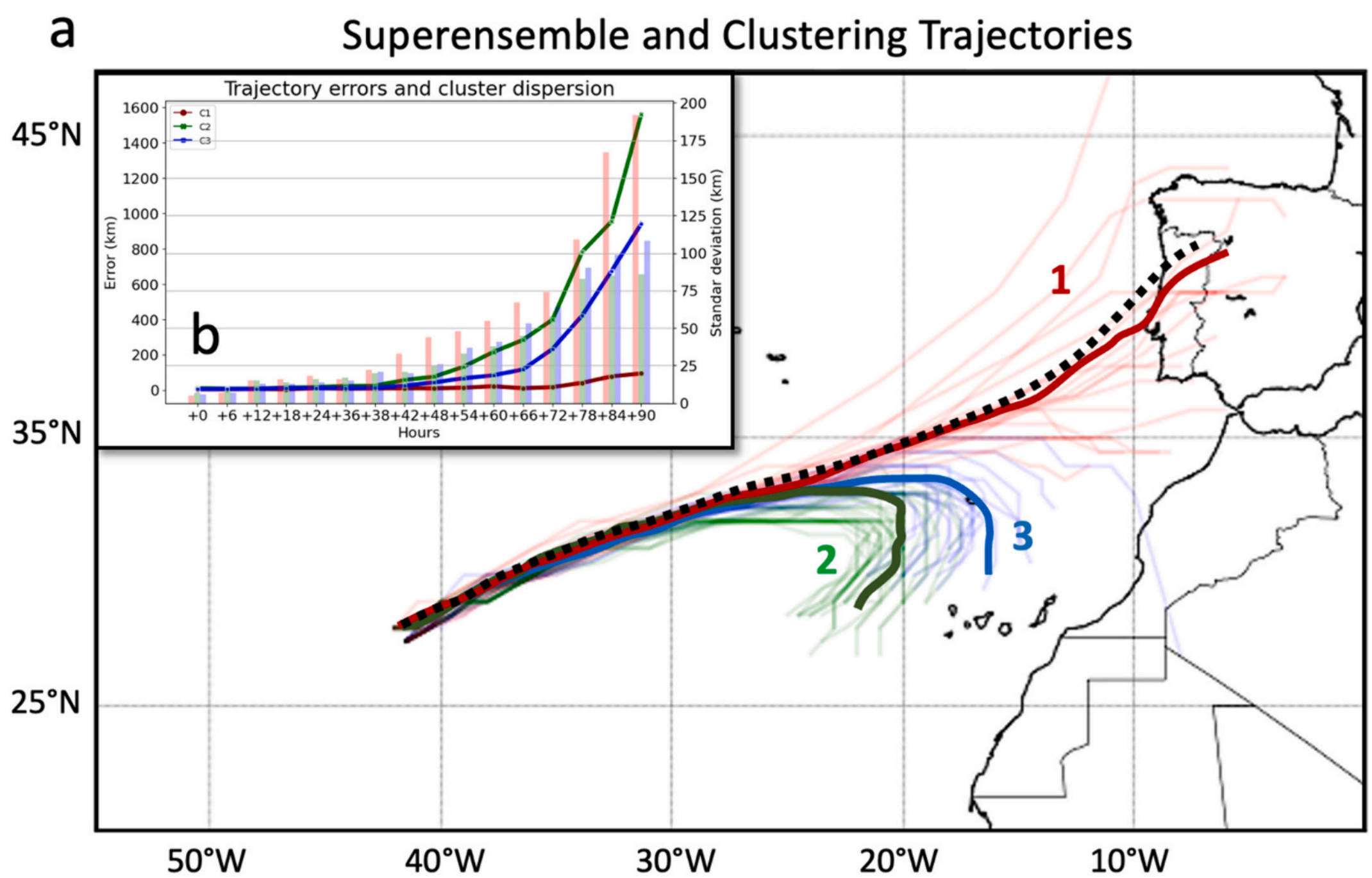

**Fig. 2.** (a) Leslie trajectories of each superensemble members, averaged path for each cluster, cluster 1 (red), cluster 2 (green), cluster 3 (blue) and best track of the National Hurricane Center, (NHC, dotted line) and (b) cluster errors with lead time with respect to the best track (lines), cluster STD (bar graphs) at 0000 UTC October 11. (For interpretation of the references to colour in this figure legend, the reader is referred to the web version of this article.)

### 3.2. Atmospheric dynamics interactions: Hurricane Michael's impact on Leslie's trajectory

Prior to October 11, Hurricane Leslie is aimlessly located in the middle of the North Atlantic Ocean (Pash and Roberts, 2019), while Hurricane Michael is undergoing rapid intensification over the warm waters of the Gulf of Mexico, reaching its maximum intensity with sustained winds of 245 km/h, which corresponds to hurricane category 4 (Beven II et al., 2019). At 0000 UTC October 11, Leslie is located at 28° N, 42° W with sustained winds of 126 km/h (category 1 hurricane, Pash and Roberts, 2019) and Michael at 31.5° N, 84.5° W is rapidly weakening after landing with sustained winds of 144 km/h (Beven II et al., 2019, Fig. 3a).

To study the propagation and modification of Rossby waves, the $PV_\theta$ concept is widely used, mainly due to its conservation property in the absence of diabatic heating and friction (Hoskins et al., 1985; Huo et al., 1999; Brennan et al., 2008). In addition, the $PV_\theta$ follows the principle of invertibility, that is, other meteorological fields can be inferred knowing the $PV_\theta$ distribution and the potential temperature (Teubler and Riemer, 2016; Miglietta et al., 2017). The moist processes and $PV_\theta$ advection by divergent wind play a prominent role in a modification of Rossby waves (Grams et al., 2015; Sánchez et al., 2020). Similarly, under situations of strong convection and diabatic heating, as in TCs, small variations in $PV_\theta$ values can propagate rapidly downstream increasing forecast errors. (Glatt and Wirth, 2014; Teubler and Riemer, 2016).

According to Hoskins et al. (1985) and used in Keller et al. (2019) and Rivière et al. (2012), cyclonic (anticyclonic) $PV_\theta$ anomalies are located on upper-level troughs (ridge). Although $PV_\theta$ anomalies are not considered in this study, differences between cluster 1 and 2 (Fig. 3) are used to identify the position and shape of both ridges located northern Michael and the downstream trough in each of the clusters. In this way, the positive (negative) differences in western (eastern) Michael area are more evident. The $PV_\theta$ composite values in cluster 1 are higher (lower) to western (eastern) Michael (Fig. 3a), with statistically significant differences between these clusters. In this context, López-Reyes et al. (2023) found that even slight differences in the initial conditions of the IFS and GFS global models can rapidly propagate downstream, leading to significant alterations in the predicted trajectory of Leslie.

In the first 12 h after initialization, the ridge-trough configuration is not clearly identified (Fig. 3b). However, using potential vorticity units ($1PVU = 10^{-6} K\, kg^{-1} m^2 s^{-1}$), the $PV_\theta$ differences around Michael become more evident between clusters, with positive values around 5 $PVU$ (negative, slightly less than 5 $PVU$) northwestern (southern) of Michael. These $PV_\theta$ positive differences over Michael could be contributing to higher ridge amplification in cluster 1 compared to cluster 2 (Fig. 3b-d), due to latent heat release processes and $PV_\theta$ destruction at upper-level associated to Michael (Fig. 3b-d), as suggested by Berman and Torn (2019) and Keller et al. (2019) in similar studies. Additionally, slight negative $PV_\theta$ differences located over Leslie show lower $PV_\theta$ composite values over Leslie in cluster 1 (Fig. 3b) indicates increased relative cyclonic vorticity ($\xi_\theta$) in cluster 1 associated with the downstream trough. In the following time steps (Fig. 3c-f), several effects are emerging: 1) A greater deepening downstream of the trough north of Leslie in cluster 1; 2) positive $PV_\theta$ differences between Michael and the ridge edge suggest a broader and more robust ridge in cluster 1; and 3) changes in the shape and propagation speed of the downstream trough. The negative (positive) differences (east) west of Leslie (Fig. 3d) indicate a slower translational speed of the ridge-trough system towards the east in cluster 1 compared to cluster 2. These differences become more evident at +60 h (Fig. 3f), when Leslie's paths in clusters 1 and 2 are completely different (Fig. 2a), indicating an influence from the upstream atmospheric pattern. In the same way, negative composite values of $PV_\theta$ extended to the north and south of the 330 K isentrope between Michael and Leslie are associated with greater intensity, elongation, and displacement to the west of the trough of cluster 1 compared to that of cluster 2 (Fig. 3e-f).

Concerning $PV_\theta$ differences, the ridge amplitude traced by the 330 K isentrope in cluster 1 is slightly larger than the one in cluster 2, 24 h after the initialization (Fig. 4a). Downstream, a slight shift in the 330 K isentrope is observed between both clusters; the trough of cluster 1 slightly develops further west (around 45°N, 50°W in red line, Fig. 4a). In the following time steps (Fig. 4b-d), as Michael displaces the 330 K isentrope, this differences between the clusters become more notable (approximately 10 *PVU* to the western with statistical significance and − 7 *PVU* northeastern Michael without statistical significance), with a maximum difference between 45° - 50° N and 50° - 70° W (Fig. 4e).

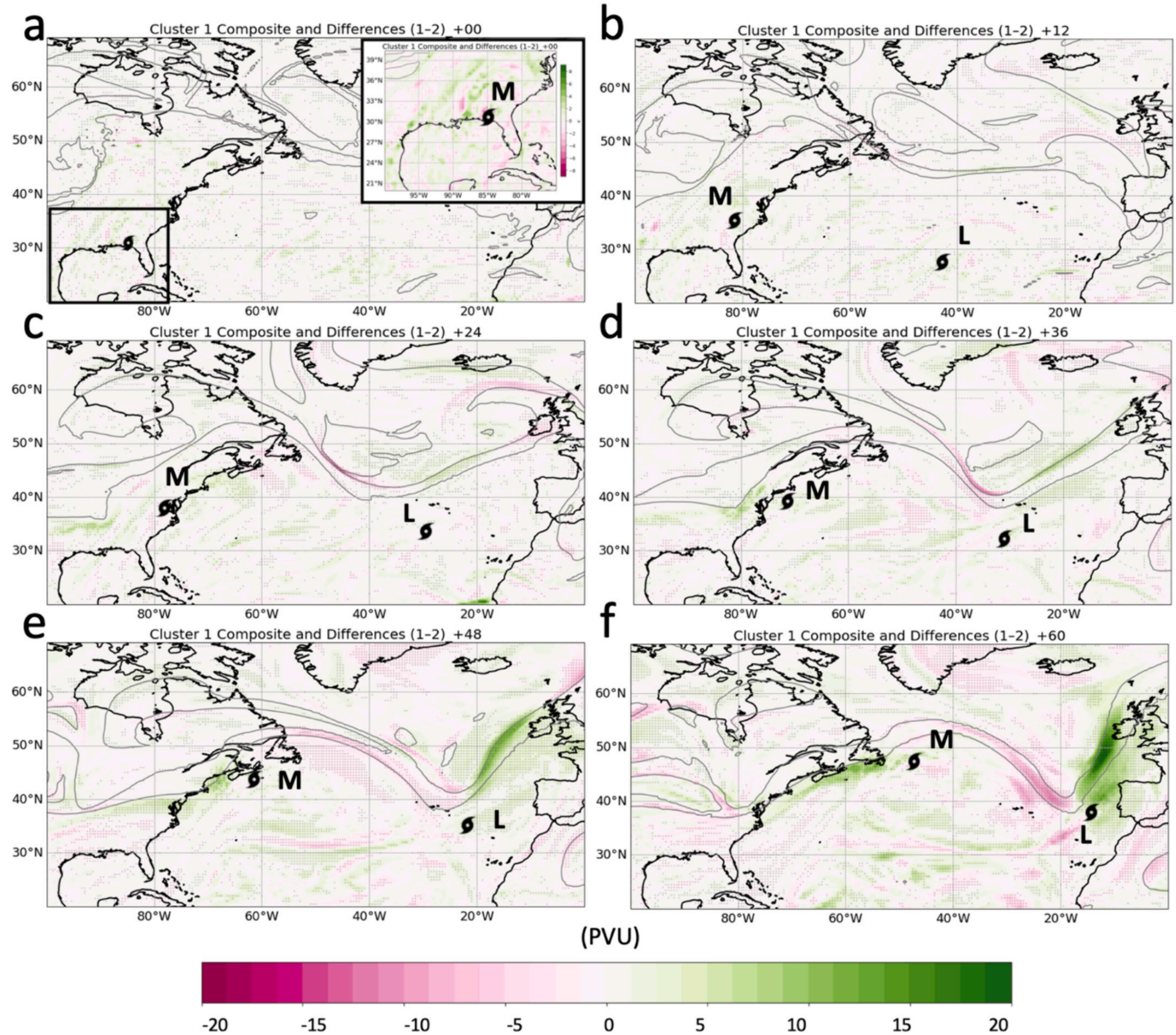

**Fig. 3.** Composite differences (cluster 1 – cluster 2) of $PV_\theta$ at 320 K (shaded; *PVU*), statistically significant differences ($\alpha = 0.05$) are shown in dotted areas on the map, $PV_\theta$ composite of cluster 1 at 320 K (contours) at: (a) +00, (b) +12, (c) +24, (d) +36, (e) +48 and (f) +60 h, and positions of Michael (M) and Leslie (L) according to NHC.

On the other hand, an elongated trough and displaced to the west is displayed in Fig. 4d compared to Fig. 4c and clearly more visible in cluster 1 ($PV_\theta$ differences around ±15 *PVU* to the eastern and western of trough axis). This is consistent with the differences in the amplitude of the upstream ridge, that is, cluster 1 (cluster 2) with greater (smaller) amplitude propagates the differences downstream, amplifying more (less) the trough. These differences in the shape and position of the trough interacted with Leslie, conditioning its trajectory (Fig. 1c). The trough of cluster 1 (red line in Fig. 4d and e) is able to interact with Leslie, steering it towards the Iberian Peninsula, just as it actually happened, according to observations. On the contrary, the trough of cluster 2, being northern (green line in Fig. 4d, e), does not interact with Leslie. In fact, cluster 2 embeds Leslie within the Azores anticyclone circulation, recurving its trajectory towards the south (Fig. 2a) instead of the best track observed.

Just like in the above-mentioned isentropic approach, the amplitude and position of ridge-trough pattern are also identified using composites of geopotential heights at 300 hPa of both clusters (Fig. S2 in Supporting Information S1). Again, small differences in an upstream ridge amplitude and position rapidly grow and change the shape of the downstream atmospheric flow (Berman and Torn, 2019). These results point towards the role of Michael in downstream predictability, and therefore its possible influence in Leslie's trajectory forecast.

To investigate how Michael may have influenced Leslie's trajectory, we focus on what occurs upstream, specifically in Michael's proximity to the ridge. To do this, the irrotational wind is derived at 300 hPa. In agreement with Berman and Torn (2019) and Teubler and Riemer (2016), the ridge amplitude modulation and stagnation are more prominently influenced by upper-tropospheric irrotational wind than by direct diabatic $PV_\theta$ modification (Eq. (4)). In our case, $\left|\vec{V}_{irr}\right|$ at 300 hPa is higher in cluster 1 at all time steps (boxplots in Fig. 5a-f and Fig. S3 in Supporting Information S1), consistent with the higher intensity of Hurricane Michael in cluster 1 (Fig. S4 in Supporting Information S1). At +12, +24 and + 36 h (Fig. 5a, c, e, and Fig. S3 in Supporting Information S1) positive (negative) values of $\left|\vec{V}_{irr}\right|$ are clearer in the northern and northwestern (northeastern) sector of Michael's position showed in green contour in Fig. 5a, c and e (purple contour in Fig. 5a, c and e). Furthermore, based on Eq. (4, the large convection associated with Michael (latent heat release) is possibly related to high values of $PV_\theta$ northwest of Michael (Fig. 3b-d); this can destroy cyclonic vorticity ($\dot{\xi}_\theta < 0$) at upper-levels, amplifying the ridge. In this way, the $\left|\vec{V}_{irr}\right|$ and $PV_\theta$ differences between both clusters, predominantly positive (negative) in western (eastern) of the dorsal axis, promote a change in the shape and intensity of the jet stream (Fig. 5a, c, e and Fig. 6a-c) that could be responsible for the differences in the position and amplitude of the downstream trough, and therefore, for the interaction with Leslie and the uncertainty in its trajectory.

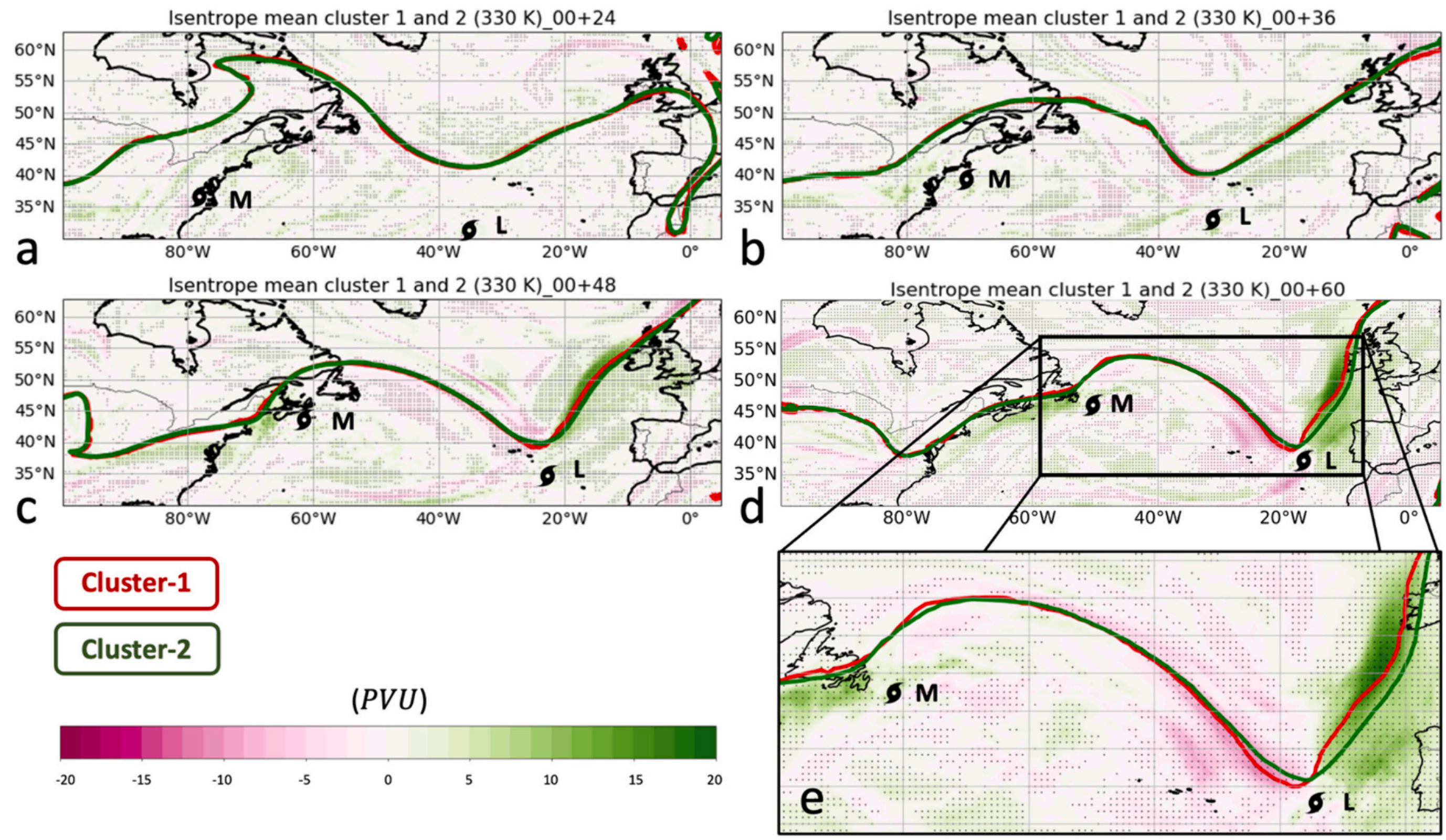

**Fig. 4.** As in Fig. 2, except for the mean 330 K isentrope of cluster 1 (red contour) and cluster 2 (green contour) at: (a) +24, (b) +36, (c) +48, (d) +60 h and (e) just like (b) with zoom. (For interpretation of the references to colour in this figure legend, the reader is referred to the web version of this article.)

The higher composite $\left|\vec{V}_{irr}\right|$ values in cluster 1, located to the northwest and north (Fig. 5a, c and e), have significant dynamical implications as follows: 1) They sustain the ridge, leading to a delayed eastward movement (Fig. 4e); 2) On the synoptic scale and based on the conservation of the $PV_\theta$, the shift of the flow towards the northeast increase $f$ values and produce in a more negative $\xi_\theta$ (assuming slight changes in atmospheric stability at upper levels on the jet region), and 3) the ridge-trough flow decelerates and adopts a more southern component earlier than in cluster 2 (Fig. 4e and red contours in Fig. 6d).

Motivated by the differences found between clusters 1 and 2 upstream, and to identify the role Michael plays more clearly in ridge-trough modifications, an analysis similar to the previous one is presented below, except for the most extreme members of clusters 1 ($P_{80}$) and cluster 2 ($P_{20}$). The extreme members are shown in Fig. 5 in Supporting Information S1. Concerning the $PV_\theta$ composite differences, the ridge-trough traced by the 330 K isentrope is shown in Fig. 7. A slightly more delayed dorsal and especially wide is observed in $P_{80}$of cluster 1. On the other hand, trough of $P_{80}$ cluster 1 downstream deepens further to the south and before than $P_{20}$ cluster 2 (close to the Leslie position) and exhibits greater curvature associated with greater $\xi_\theta$. Then, we have stronger evidence about the influence of the upstream ridge that Michael is interacting with on the uncertainty of Leslie's trajectory.

Concerning clusters 1 ($P_{80}$) and cluster 2 ($P_{20}$) of $\vec{V}_{irr}$ and jet stream, in Figs. 8 and 9 are represented, respectively. During the first +12 h the $\left|\vec{V}_{irr}\right|$ differences of $P_{80}$ cluster 1 and $P_{20}$ cluster 2 (Fig. 8a) maintain very similar behavior respect to the cluster differences around Michael, approximately $+2\ ms^{-1}$ (Fig. 8a). However, at +24 and + 36 h the $\left|\vec{V}_{irr}\right|$ in $P_{80}$ cluster 1 is more intense than the cluster 1 mean (Fig. 8b, c). Again, stronger positive (negative) differences predominate northwest–north (northeastern) of Michael's position $+3\ ms^{-1}$ ($-2\ ms^{-1}$) in green and purple contours in Fig. 8b and c. $\left|\vec{V}_{irr}\right|$ greater values $P_{80}$ cluster 1 are consistent with the more northerly propagation of the 330 K isentrope at +60 h, as shown in Fig. 7.

As might be suspected, the clusters extremes mean differences are capable of evidencing the source of uncertainty. As mentioned before, a greater $\vec{V}_{irr}$ associated with a more intense Michael is able to slow down and amplify the ridge. This situation modifies the development of the downstream jet stream as shown in Fig. 9. When comparing the differences in the shape and position of the jet stream, it is observed that the ridge-trough flow in $P_{80}$ cluster 1 is delayed, adopts a more southern component than in cluster 2 (Fig. 9). Likewise, positive differences southern of the jet streak (Fig. 9d) confirm a greater depth of the trough in $P_{80}$ cluster 1 which may have facilitated Leslie's immersion in the flow.

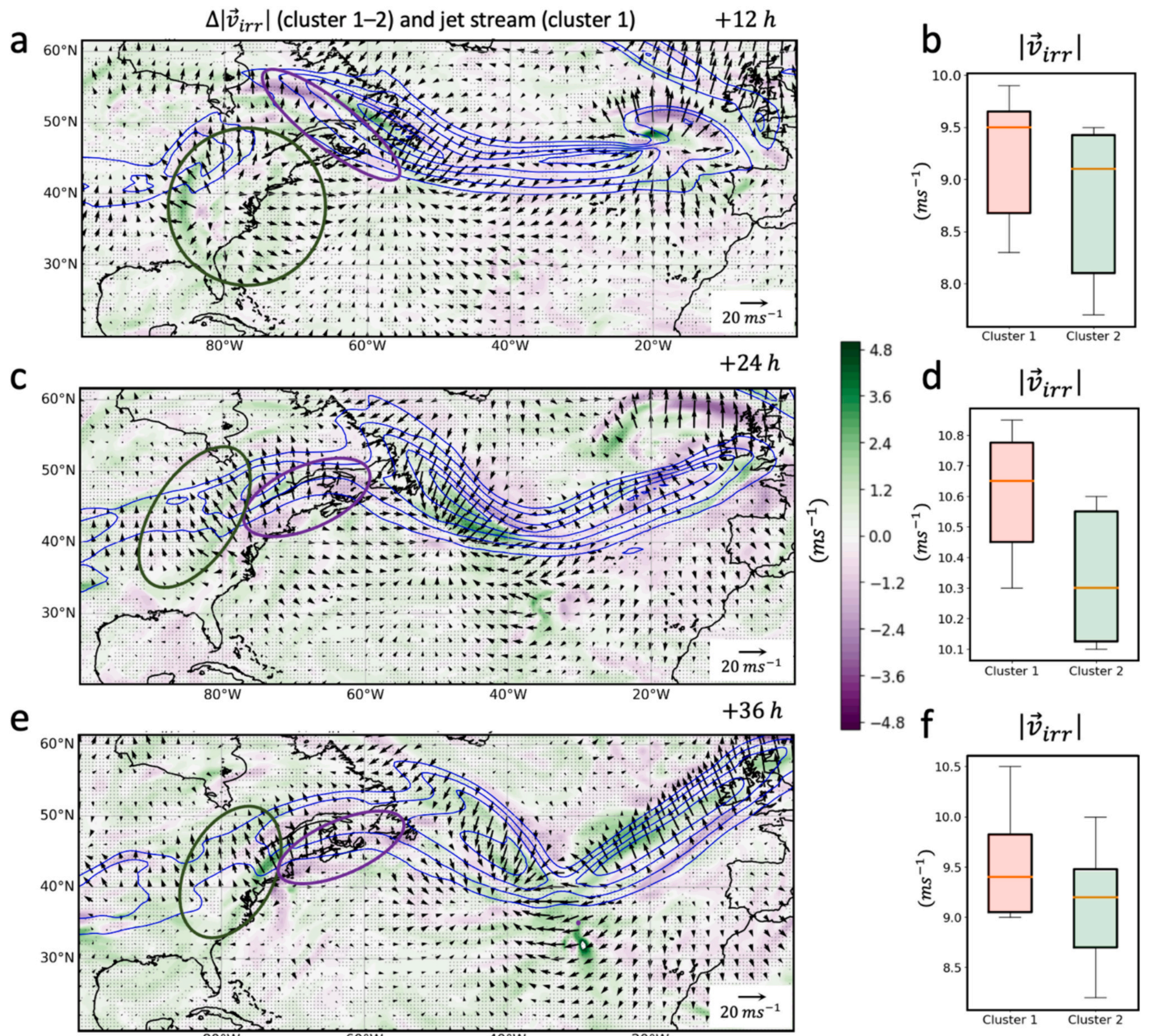

**Fig. 5.** Irrotational wind (vectors; $ms^{-1}$), $\left|\vec{V}_{irr}\right|$ composite differences (cluster 1 – cluster 2; shaded; $ms^{-1}$) with statistical significance (dot) and jet stream of cluster 1 (blue contours between 40 and 110 $ms^{-1}$ every 10 $ms^{-1}$) at 300 $hPa$ at: (a) +12, (c) +24 and (e) +36 h. $\left|\vec{V}_{irr}\right|$ values ($ms^{-1}$) of each cluster in box plots at: (b) +12, (d) +24 and (f) +36 h. (For interpretation of the references to colour in this figure legend, the reader is referred to the web version of this article.)

For de above and based on the Michael's and Leslie's positions (Fig. S1 in Supporting Information S1), as well as the patterns of the $\left|\vec{V}_{irr}\right|$ (Fig. 5e and 8c), $PV_\theta$ (Figs. 4d and 7), and jet stream configuration (Fig. 6d and 9d), a conceptual scheme of the Michael-Leslie situation is presented in Fig. 10. The patterns of $\left|\vec{V}_{irr}\right|$ and $PV_\theta$ are responsible for modifying the ridge-trough shape above Michael. In cluster 1 and $P_{80}$ cluster 1, the greater $\left|\vec{V}_{irr}\right|$ values to the western and northwest of the ridge cause the ridge-trough system to be delayed and amplified slightly more than in cluster 2 and $P_{20}$ cluster 2 (Fig. 4e and 7). Furthermore, the greater $\left|\vec{V}_{irr}\right|$ values of cluster 1 and $P_{80}$ cluster 1 in the northern and northwestern (Figs. 5 and 8) caused the jet stream to be delayed and displaced towards the south. This phenomenon contributed to the amplification of the trough downstream, which effectively captured Hurricane Leslie and redirected its path towards the Iberian Peninsula, consistent with the observed best track data (Fig. 2a). In cluster 2 and $P_{20}$ cluster 2, Leslie does not interact in the trough downstream, being to the west and influenced by the circulation of the Azores anticyclone, the system curves its path southward (Fig. 2a).

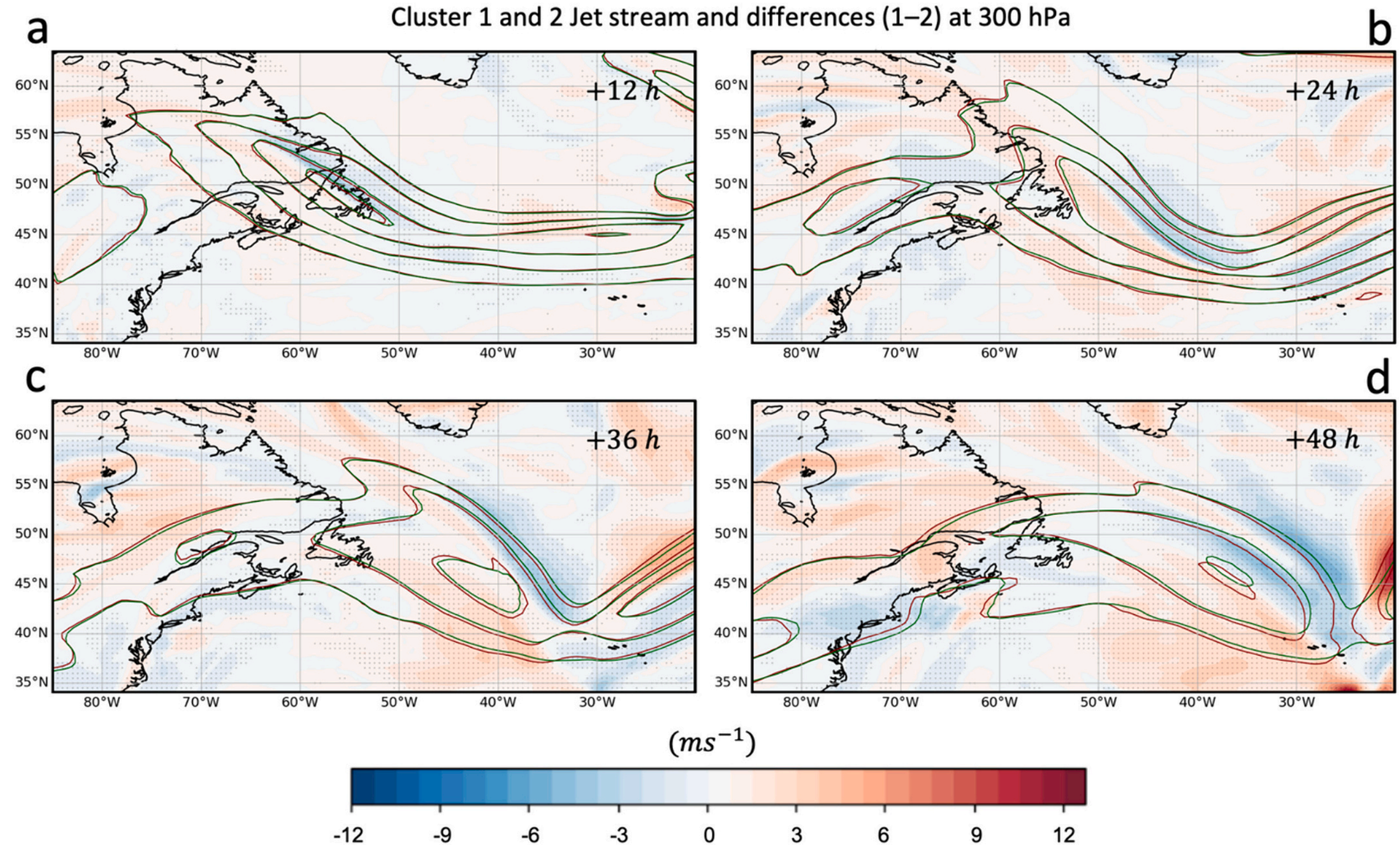

**Fig. 6.** Jet stream for cluster 1 (red contour) and 2 (green contour) between 40 and 110 $ms^{-1}$ every 20 $ms^{-1}$. Wind velocity differences cluster 1 – cluster 2 (shaded; $ms^{-1}$) with statistical significance (dot) at 300 hPa at (a) +12, (b) +24, (c) +36 h and (d) +48 h. (For interpretation of the references to colour in this figure legend, the reader is referred to the web version of this article.)

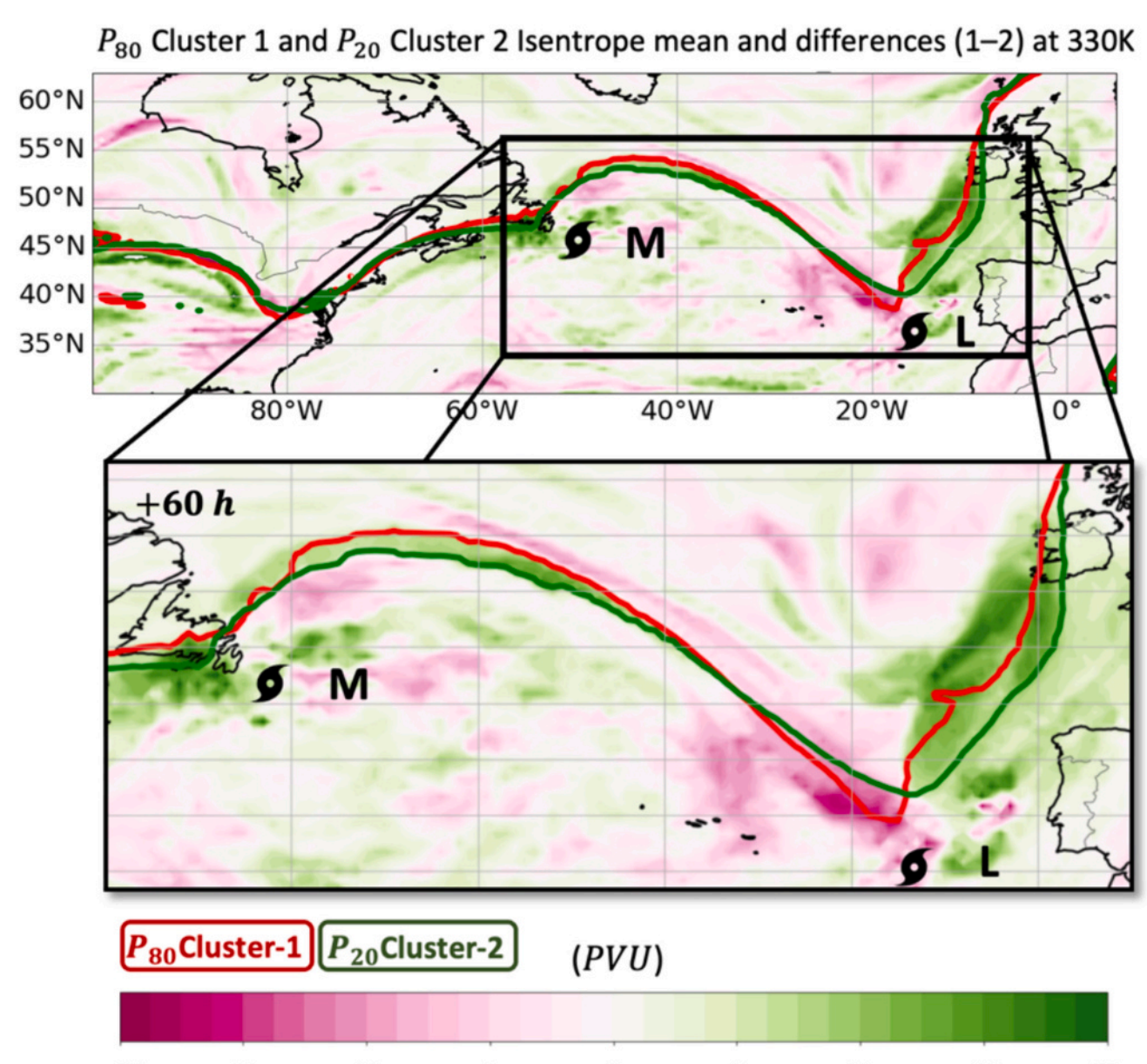

**Fig. 7.** As in Fig. 4d but for $P_{80}$of cluster 1 (red contour) and $P_{20}$ of cluster 2 (green contour). Since there are few members, statistical significance is not shown. (For interpretation of the references to colour in this figure legend, the reader is referred to the web version of this article.)

## 4. Summary and conclusions

This study evaluates the uncertainty in the track forecasts of Hurricane Leslie during its ET phase at the time of the greatest dispersion. 70 perturbed members from EPS and GEFS are used, and a clustering method based on finite mixture model is applied to identify the upstream role of Hurricane Michael on the downstream Leslie trajectory. Differences between the most extreme groups (cluster 1 and 2) and for the subset formed by the $P_{80}$ cluster 1 and $P_{20}$ cluster 2 are considered to explain the great uncertainty in Leslie's trajectory.

Although cluster 1 shows great dispersion in its members, the averaged trajectory reproduces quite well the best track for all the time steps, taking Leslie towards the Iberian Peninsula. On the contrary, in cluster 2 all members recurve southward the Leslie's trajectory transiting western the Canary Islands. From the beginning of initialization (October 11, 2018, at 0000 UTC), positive (negative) differences are already shown in the $PV_\theta$ to the western (eastern) Michael's position in all members in the superensemble. These differences between both clusters are clearly shown over the ridge-trough pattern, especially in the trough after +24 h, reaching ±15 *PVU* around the trough axis with higher southern negative values. This indicates that the trough in cluster 1 has greater amplitude and a more slowly easterly displacement in comparison with cluster 2 (even more evident when taking the $P_{80}$ cluster 1 and the $P_{20}$ cluster 2). Upstream, the $PV_\theta$ and $\left|\overrightarrow{V}_{irr}\right|$ associated with Michael, exhibit notable differences. On the one hand, in cluster 1 (with Michael more intense) higher values of $PV_\theta$ and $\left|\overrightarrow{V}_{irr}\right|$ are presented (probably related to Michael's heat release processes and $PV_\theta$ destruction), especially to the western and northern of the ridge axis. This favors the ridge to grow towards the west and slow down its movement east with respect to cluster 2 (more evident when taking the $P_{80}$ cluster 1 and the $P_{20}$ cluster 2), as shown in the conceptual scheme (Fig. 10). The shape and propagation of the jet stream of cluster 1 (delayed and broader downstream) is consistent with the $\left|\overrightarrow{V}_{irr}\right|$ higher values and spatial distribution in cluster 1. This is the key to connecting Hurricane Michael to the high uncertainty associated with the track of Hurricane Leslie. The differences in the shape and position of the ridge, depicted by both clusters, propagate downstream through the jet stream and drastically modifying the amplitude and position of the trough near to Leslie. In the case of cluster 1, the southernmost and delayed trough allows Leslie to be embedded in the atmospheric flow maintaining a northeastward trajectory. On the other hand, the less elongated trough with faster movement to the east does not draw in Leslie, so being over the Azores anticyclone it recurves towards the south.

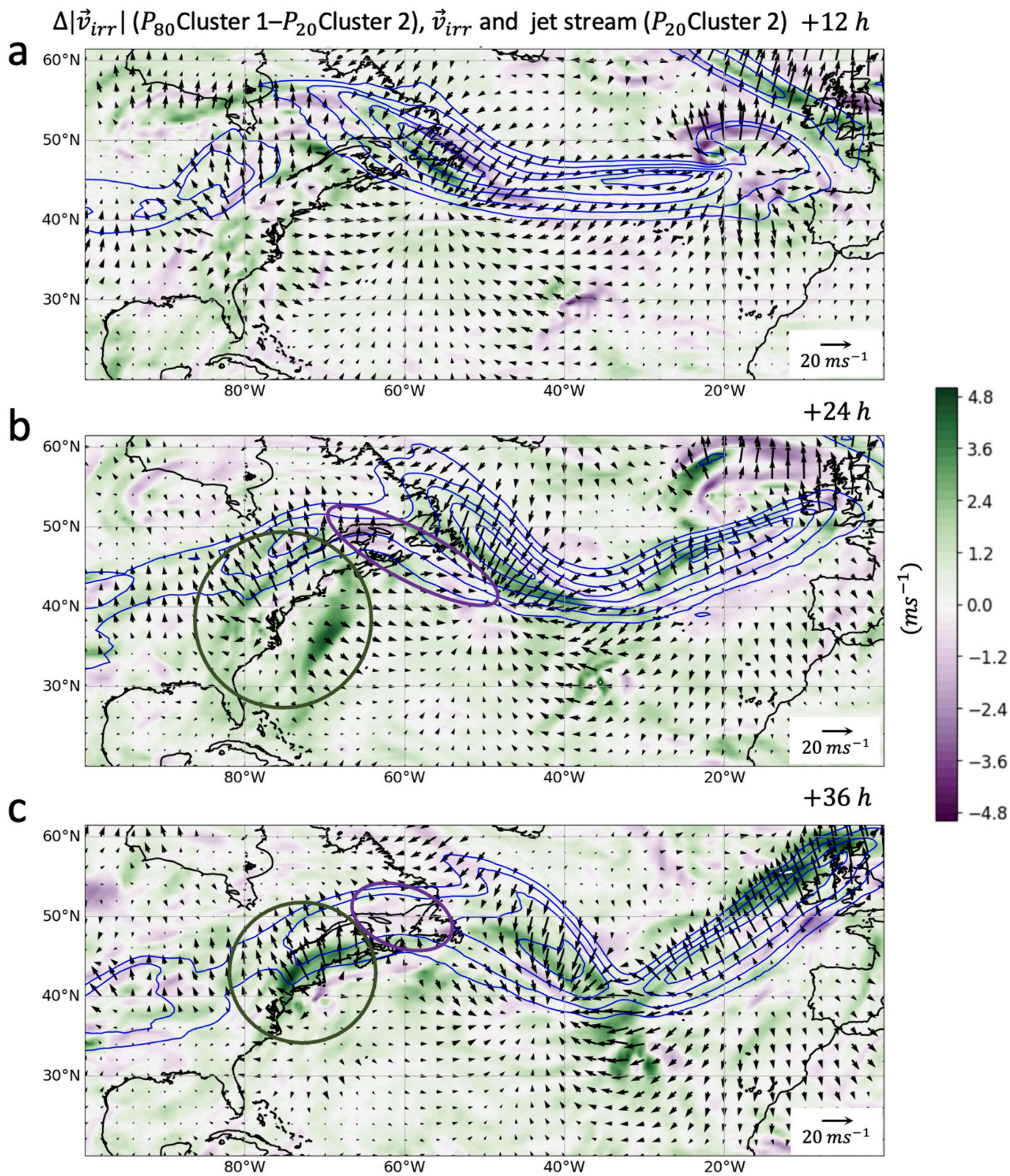

**Fig. 8.** As in Fig. 5 but for $P_{80}$of cluster 1 – $P_{20}$ of cluster 2 and $P_{80}$of cluster 1 jet stream.

This study shows how slight changes in the shape (intensity) and position of an upstream ridge, under intense diabatic processes related to hurricane convection and the behavior of irrotational wind at high levels, can quickly alter the atmospheric pattern, significantly increasing uncertainty in downstream weather forecasting. This result is especially relevant for operational meteorologist since, the proposed conceptual scheme can be used to evaluate scenarios of high uncertainty in the trajectory forecast under situations where there are interactions between two cyclones with the jet stream.

## Open research

The tracking data for Leslie can be found in López-Reyes (2024). Additionally, atmospheric data sets can be accessed through the TIGGE database, hosted by ECMWF, at https://confluence.ecmwf.int/display/TIGGE.

## CRediT authorship contribution statement

**M. López-Reyes:** Writing – original draft, Software, Methodology, Investigation, Formal analysis, Data curation, Conceptualization. **J.J.**

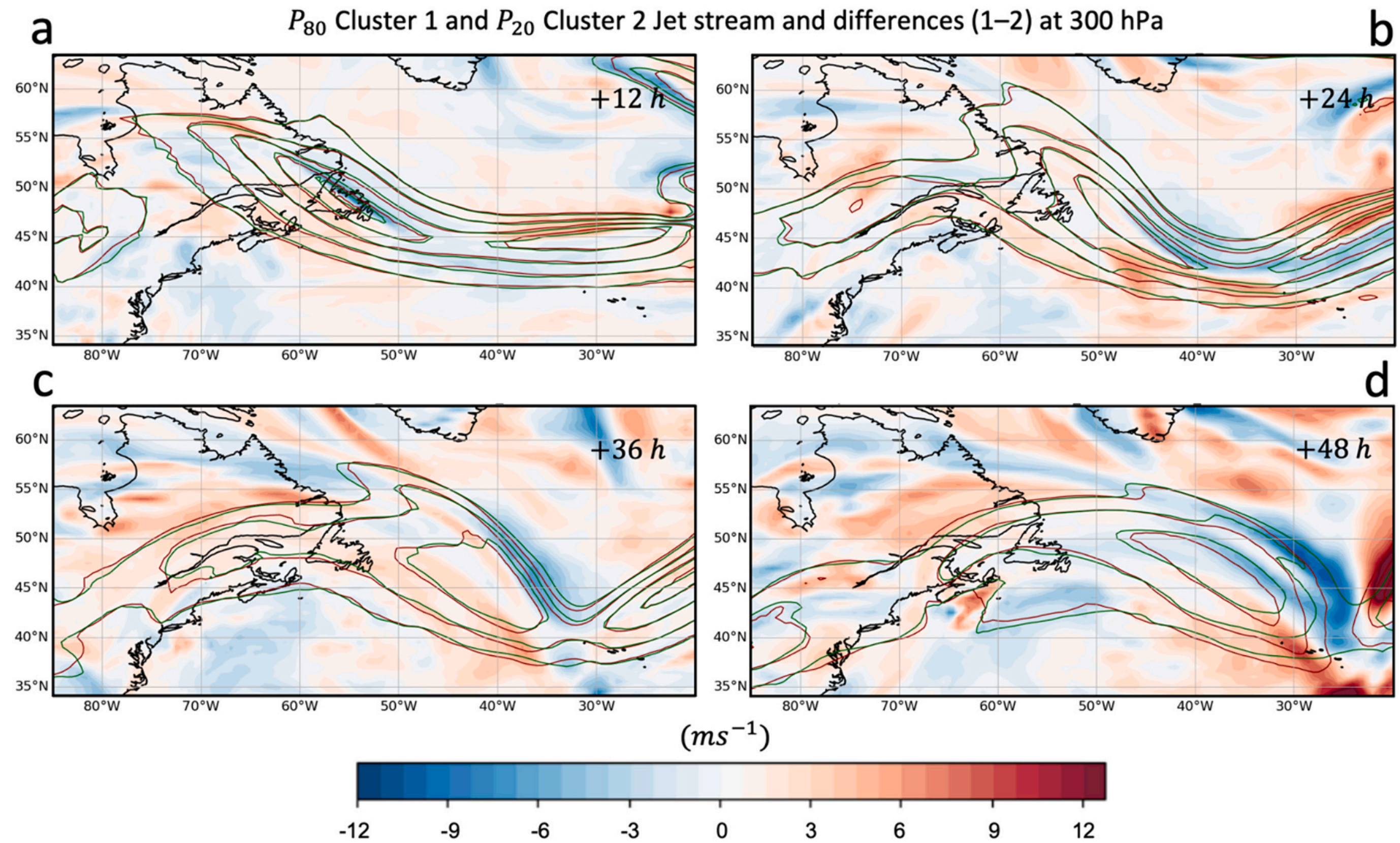

**Fig. 9.** As in Fig. 6 but for $P_{80}$of cluster 1 (red contour) and $P_{20}$ of cluster 2 (green contour). (For interpretation of the references to colour in this figure legend, the reader is referred to the web version of this article.)

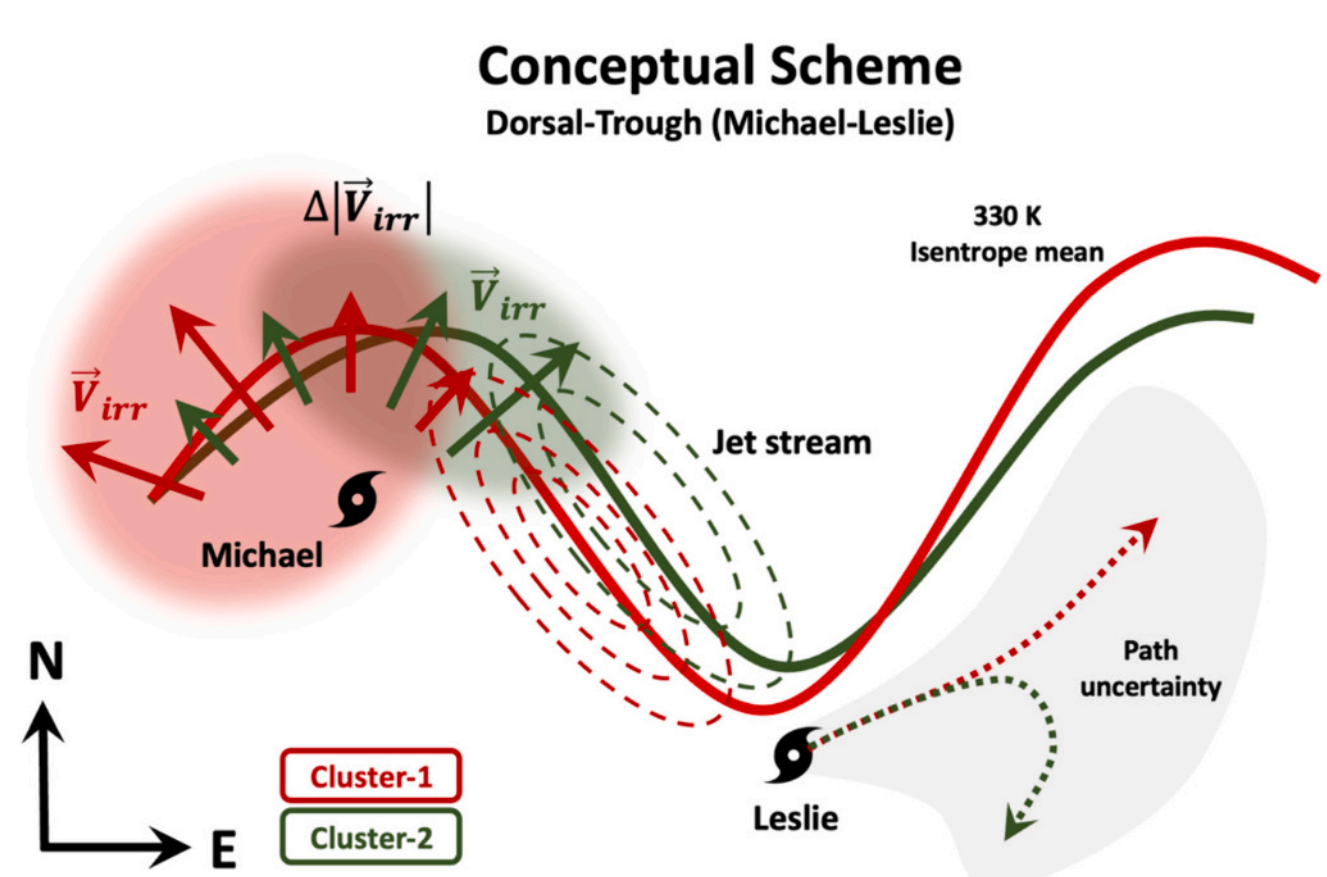

**Fig. 10.** Dorsal-trough conceptual scheme based in $\left|\vec{V}_{irr}\right|$, $PV_\theta$ and jet stream patters, cluster 1 and $P_{80}$ cluster 1 (red) and cluster 2 and $P_{20}$ cluster 2 (green). Continues lines are isentrope mean at 330 K of both clusters, vectors indicate the direction and magnitude of the irrotational wind associated with Hurricane Michael. $\left|\vec{V}_{irr}\right|$ differences between clusters are shown in shaded areas; dashed lines indicate the position and intensity of the jet stream; dotted lines indicate the Leslie trajectories associated with each cluster and grey shaded region represent the path uncertainty. (For interpretation of the references to colour in this figure legend, the reader is referred to the web version of this article.)

**Gónzalez-Alemán:** Writing – review & editing, Supervision, Methodology, Investigation, Formal analysis, Conceptualization. **C. Calvo-Sancho:** Writing – review & editing, Software, Data curation. **P. Bolgiani:** Writing – review & editing, Supervision, Investigation. **M. Sastre:** Writing – review & editing, Supervision, Investigation, Formal analysis. **M.L. Martín:** Writing – review & editing, Supervision, Resources, Project administration, Methodology, Investigation, Funding acquisition, Formal analysis, Conceptualization.

## Declaration of competing interest

The authors declare no conflicts of interest relevant to this study.

## Data availability

Data will be made available on request.

## Acknowledgments

This work was partially supported by the research project PID2019-105306RB-I00/AEI/10.13039/501100011033 (IBERCANES), and the two ECMWF Special Projects (SPESMART and SPESVALE). Mauricio López-Reyes extends his sincere gratitude to Professor Héctor Ulloa-Godínez from the Institute of Astronomy and Meteorology at the University of Guadalajara for his invaluable support. He also acknowledges Instituto Frontera A.C. for their partial funding of this work. C. Calvo-Sancho acknowledges the grant awarded by the Spanish Ministry of Science and Innovation - FPI Program (PRE2020-092343). We would like to thank Instituto Frontera A.C. for financial support to carry out the work.

## Appendix A. Supplementary data

Supplementary data to this article can be found online at https://doi.org/10.1016/j.atmosres.2024.107697.